\documentstyle[aps,epsf]{revtex}  

\newcommand{\be}{\begin{equation}}
\newcommand{\ee}{\end{equation}}
\newcommand{\ben}{\begin{eqnarray}}
\newcommand{\een}{\end{eqnarray}}

\def\simgt{\rlap{\lower 3.5 pt\hbox{$\mathchar \sim$}}\raise 1pt \hbox {$>$}}
\def\simlt{\rlap{\lower 3.5 pt\hbox{$\mathchar \sim$}}\raise 1pt \hbox {$<$}}

\begin{document}        

\baselineskip 14pt
\title{CP-PACS results for light hadron spectrum in quenched 
and two-flavor full QCD}
\author{Yoshinobu Kuramashi 
\protect{\thanks{On leave from Institute of Particle and Nuclear Studies,
High Energy Accelerator Research Organization(KEK),
Tsukuba, Ibaraki 305-0801, Japan}}
for the CP-PACS Collaboration
\protect{\thanks{CP-PACS Collaboration: S.~Aoki, G.~Boyd, R.~Burkhalter,
S.~Ejiri, M.~Fukugita, S.~Hashimoto, Y.~Iwasaki, K.~Kanaya, 
T.~Kaneko, Y.~K., K.~Nagai, M.~Okawa, H.~P.~Shanahan, A.~Ukawa, 
and T.~Yoshi{\'e}}}}
\address{Department of Physics, Washington University, 
St.~Louis, Missouri 63130}
%
\maketitle              

\begin{abstract}        
We present a summary of our results for the light hadron spectrum
in quenched lattice QCD and preliminary ones in two-flavor full QCD.
For the quenched hadron spectrum 
we find that the mass formulae predicted by quenched chiral 
perturbation theory(QChPT) give a good description of 
our results. Employing the QChPT mass formulae 
for chiral extrapolations
we conclude that the mass spectrum in the continuum limit 
deviates unambiguously and systematically from experiment.
For our two-flavor full QCD results we focus on the dynamical 
quark effects on the light hadron spectrum. 

\
\end{abstract}   	

\section{Introduction}               



Numerical studies of QCD on the lattice
provide us a quantitative understanding of the dynamics 
of strong interactions.
One of the main goals is to derive the hadron mass spectrum
from the first principle.
In order to give a precise prediction with lattice QCD
simulations we have to control the following 
three physical sources of
systematic errors besides the quenching error: 
(i) scaling violation, (ii) finite size effects
and (iii) chiral extrapolations. 

In the quenched approximation the GF11 collaboration 
carried out an extensive 
calculation to reduce these three systematic errors and found
that the quenched spectrum is consistent with experiment
within $5-10\%$ errors\cite{GF11}.
The aim of our work is to tell definitely how the quenched
spectrum deviates from experiment, by considerably 
diminishing the statistical and systematic uncertainties.  

On the other hand, 
in spite of much effort devoted in the full QCD spectrum study
we have not yet been able to answer a fundamental question:
to what extent the dynamical quarks affect the light hadron spectrum.  
Before embarking upon a realistic full QCD calculation in the future
we attempt to settle this question. 

In the first part of this report we present our results 
for the quenched light hadron spectrum, examining the validity
of QChPT for the chiral extrapolation.
The two-flavor full QCD results obtained so far are presented 
in the second part, where we investigate possible signs of
dynamical quark effects on the light hadron spectrum. 
More details are found in Ref\cite{latt98}.

\section{Quenched light hadron spectrum with the Wilson quark action}


\subsection{Details of numerical simulation}

\begin{table}[t]
\caption{\label{tab:para_quench} 
Simulation parameters for quenched QCD.}
\begin{tabular}{llllllllllll} 
$\beta$ & $L^3\times T$ & $a^{-1}$[GeV] & $La$[fm] & \#conf. & 
$\begin{array}{c}{\rm sweep}\\{\rm /conf.}\end{array}$ &
\multicolumn{5}{c}{$m_\pi/m_\rho$} & $\delta$ \\
\tableline 
$5.90$ &  $32^3\times 56$  & $1.934(16)$ & $3.26(3)$ & $800$ &  $200$ &
0.752(1) & 0.692(1) & 0.593(1) & 0.491(2) & 0.415(2) &
0.106(5)  \\
$6.10$ &  $40^3\times 70$  & $2.540(22)$ & $3.10(3)$ & $600$ &  $400$ &
0.751(1) & 0.684(1) & 0.581(2) & 0.474(2) & 0.394(3) &
0.103(6)  \\
$6.25$ &  $48^3\times 84$  & $3.071(34)$ & $3.08(3)$ & $420$ & $1000$ &
0.760(1) & 0.707(2) & 0.609(2) & 0.502(2) & 0.411(3) &
0.117(7)  \\
$6.47$ &  $64^3\times 112$ & $3.961(79)$ & $3.18(6)$ & $150$ & $2000$ &
0.759(2) & 0.708(3) & 0.584(3) & 0.493(4) & 0.391(4) &
0.113(13) \\
\end{tabular}
\end{table}

Our simulations are carried out with the plaquette gauge action 
and the Wilson quark action.  
To control the systematic errors
we carefully choose our run parameters, which are summarized in
Table~\ref{tab:para_quench}. We employ four $\beta$ values 
so that the lattice spacing covers the range $a\approx 0.1-0.05$fm 
to remove the scaling violation effects
by extrapolation of the data to the continuum limit. 
To avoid finite size effects we keep
the physical spatial lattice size approximately constant
at $La\approx 3$fm. A previous study showed that the finite size effects
are $2\%$ or less already at $La\approx 2$fm\cite{finiteV}.
The most subtle issue in controlling the systematic errors 
is associated with chiral extrapolations of the hadron masses. 
Although the mass formulae predicted by QChPT are considered 
to be plausible candidates for the fitting functions 
of chiral extrapolations, their validities
should be checked employing a wide range of quark masses.  
At each value of $\beta$ we choose five values of the hopping parameter
$K$ corresponding to $m_\pi/m_\rho\approx 0.75$, $0.7$, $0.6$, 
$0.5$ and $0.4$. The heaviest two values $m_\pi/m_\rho\approx 0.75$ 
and $0.7$ are taken to be around the physical strange quark mass.
In terms of these quark masses we calculate hadron masses 
both for degenerate and non-degenerate cases.

Gauge configurations are generated with the 5-hit pseudo heat-bath
algorithm incorporating the over-relaxation procedure quadruply
and employing periodic boundary condition. 
Quark propagators are solved in the Coulomb gauge both 
for the point and the smeared sources with periodic boundary
conditions imposed in all four directions. For the hadron mass 
measurement we use the smeared source quark propagators.

The physical point for the degenerate up and down quark mass is
determined by $m_\pi(135.0)$ and $m_\rho(768.4)$.
For the strange quark mass we employ $m_K(497.7)$ or $m_\phi(1019.4)$.
The lattice scale $a^{-1}$ is set with $m_\rho$.  

\subsection{Chiral extrapolations}

\begin{figure}[b]
\begin{minipage}[b]{80mm}
\centerline{\epsfxsize 80mm \epsfbox{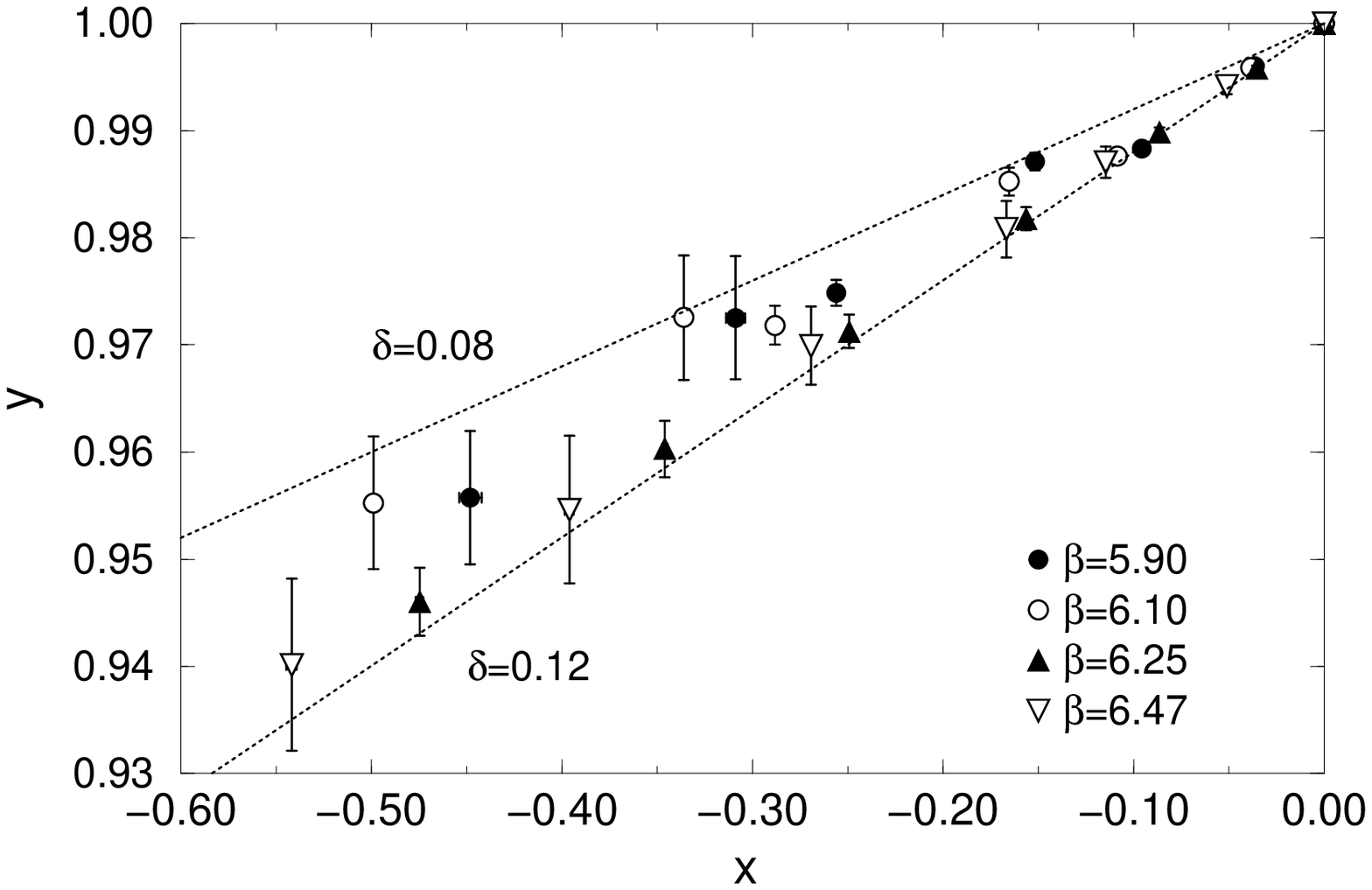}}   
\caption[]{\label{fig:delta}
Test of the quenched chiral logarithm using the PS meson masses.}
\vspace*{3.5mm}
\end{minipage}
\hspace{\fill}
\begin{minipage}[b]{80mm}
\centerline{\epsfxsize 73mm \epsfbox{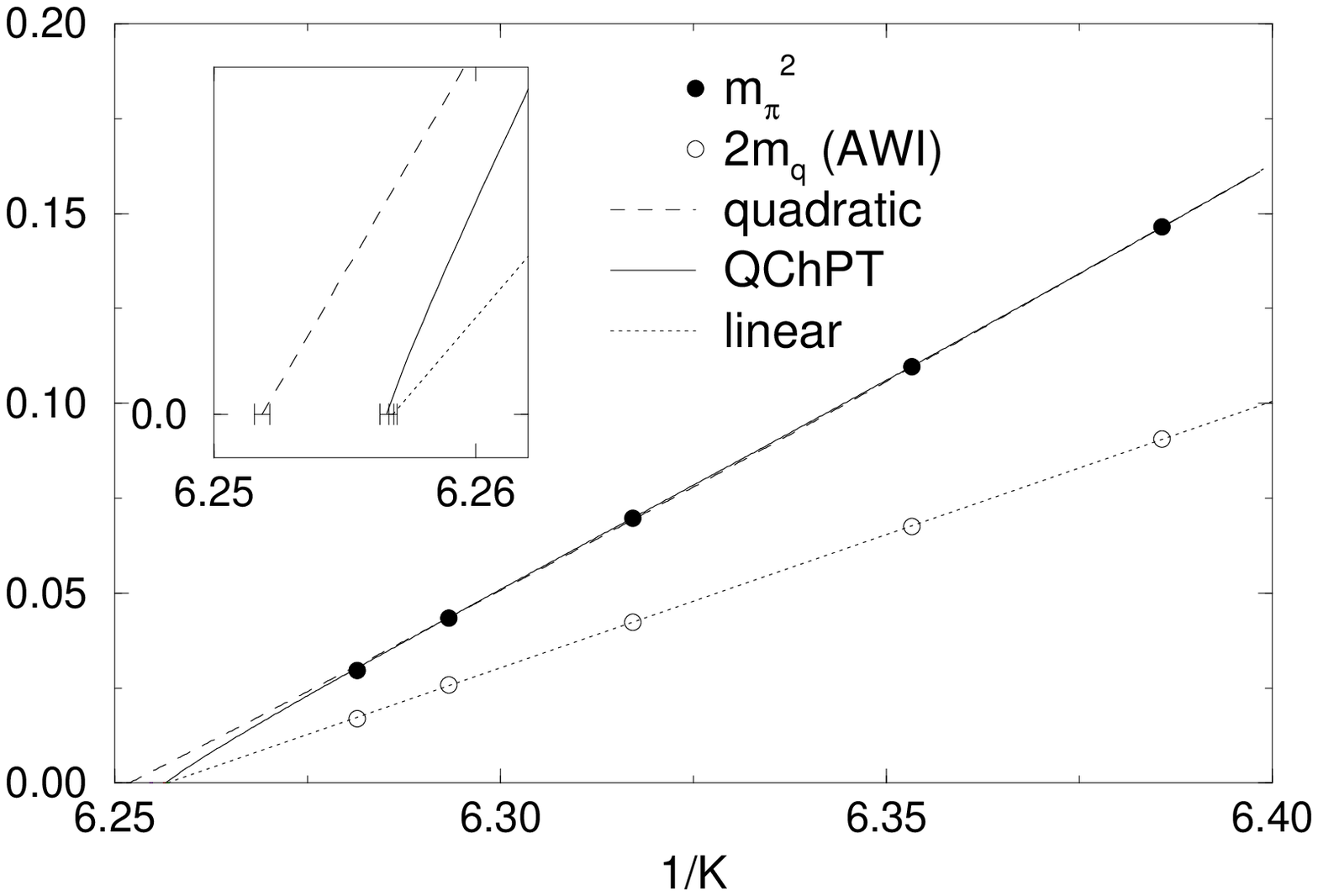}}   
\caption[]{\label{fig:ps_chl_quench}
Chiral extrapolations of the degenerate PS meson mass and the
AWI quark mass as a function of $1/K$ at $\beta=5.9$.}
\end{minipage}
\end{figure}

We first examine the validity of QChPT
investigating the presence of quenched chiral logarithm
in the pseudoscalar(PS) meson sector.
For the quark mass dependence of the PS meson mass 
$m_{PS}$ the QChPT predicts\cite{QChPT} 
\be
\frac{m_{PS}^2}{m_s+m}=A\left\{1-\delta\left[{\rm ln}
\left(\frac{2mA}{\Lambda_\chi^2}\right)
+\frac{m_s}{m_s-m}{\rm ln}\left(\frac{m_s}{m}\right)\right]\right\}
+B(m_s+m)+O((m,m_s)^2),
\label{eq:m_ps}
\ee 
where $m$ and $m_s$ are masses of two valence quarks in the PS meson.
This ratio diverges logarithmically 
toward the chiral limit due to the $\delta$ term.
To detect the contribution of 
the $\delta$ term in a direct manner we introduce
the two  variables: 
$x=2-(m_s+m){\rm ln}(m_s/m)/(m_s-m)$ and
$y=4 m m_s/(m_s+m)^2\times m_K^4/(m_\pi^2 m_\eta^2)$,
for which eq.(\ref{eq:m_ps}) leads to the relation $y=1+\delta\cdot x$.
Here $\pi$ and $\eta$ are the degenerate PS mesons with
quark mass $m$ or $m_s$ and $K$ is the non-degenerate one with
$m$ and $m_s$. It should be noted that the quark masses are
defined by an extended axial vector current Ward identity(AWI)
$\nabla_\mu A_\mu^{\rm ext}=2m_q^{\rm AWI} P$\cite{m_q^AWI},
where we are free from ambiguities originating from the
determination of the critical hopping parameter $K_c$.
Figure~\ref{fig:delta} shows the distribution of $y$ as a function 
of $x$. Our data fall in a wedge shaped with $y=1+0.08 x$ 
and $y=1+0.12 x$. 
A PS meson decay constant ratio $y=f_K^2/(f_\pi f_\eta)$, for which 
QChPT predicts $y=1-\delta/2\cdot x$, 
is another quantitative test for $\delta$.
Our data indicate that the value of $\delta$ is within 
$\delta=0.08-0.16$. 

Since we observe clear evidence 
for the existence of the quenched chiral
logarithm, we now try to fit the PS meson masses 
with the functional form of eq.(\ref{eq:m_ps}).  
In Fig.~\ref{fig:ps_chl_quench} we show a typical result
for chiral extrapolations of the degenerate
PS meson mass and the AWI quark mass.
There are two features to be remarked.
One is a good description of the AWI quark mass 
with a linear function of $1/K$, which means 
the AWI quark mass is proportional 
to the vector Ward identity(VWI) quark mass 
$m_q^{\rm VWI}=(1/K-1/K_c)/2$.
The other is a good agreement between the critical hopping parameter
$K_c$ determined from the QChPT fit of $m_{PS}^2$ 
and that from the linear fit of $m_q^{\rm AWI}$.
They are consistent within $2.5\sigma$, which should be compared
with $17\sigma$ ($12\sigma$) discrepancy for the case of
a quadratic (cubic) polynomial fit of $m_{PS}^2$.  

Our examinations for the chiral properties of the PS meson masses
strongly suggest the validity of QChPT 
with $\delta\approx 0.10(2)$. It would be legitimate to apply
the QChPT mass formulae for the
chiral extrapolation  of the vector meson masses 
and the baryon ones.  

\begin{figure}[b]
\begin{minipage}[b]{80mm}
\centerline{\epsfxsize 73mm \epsfbox{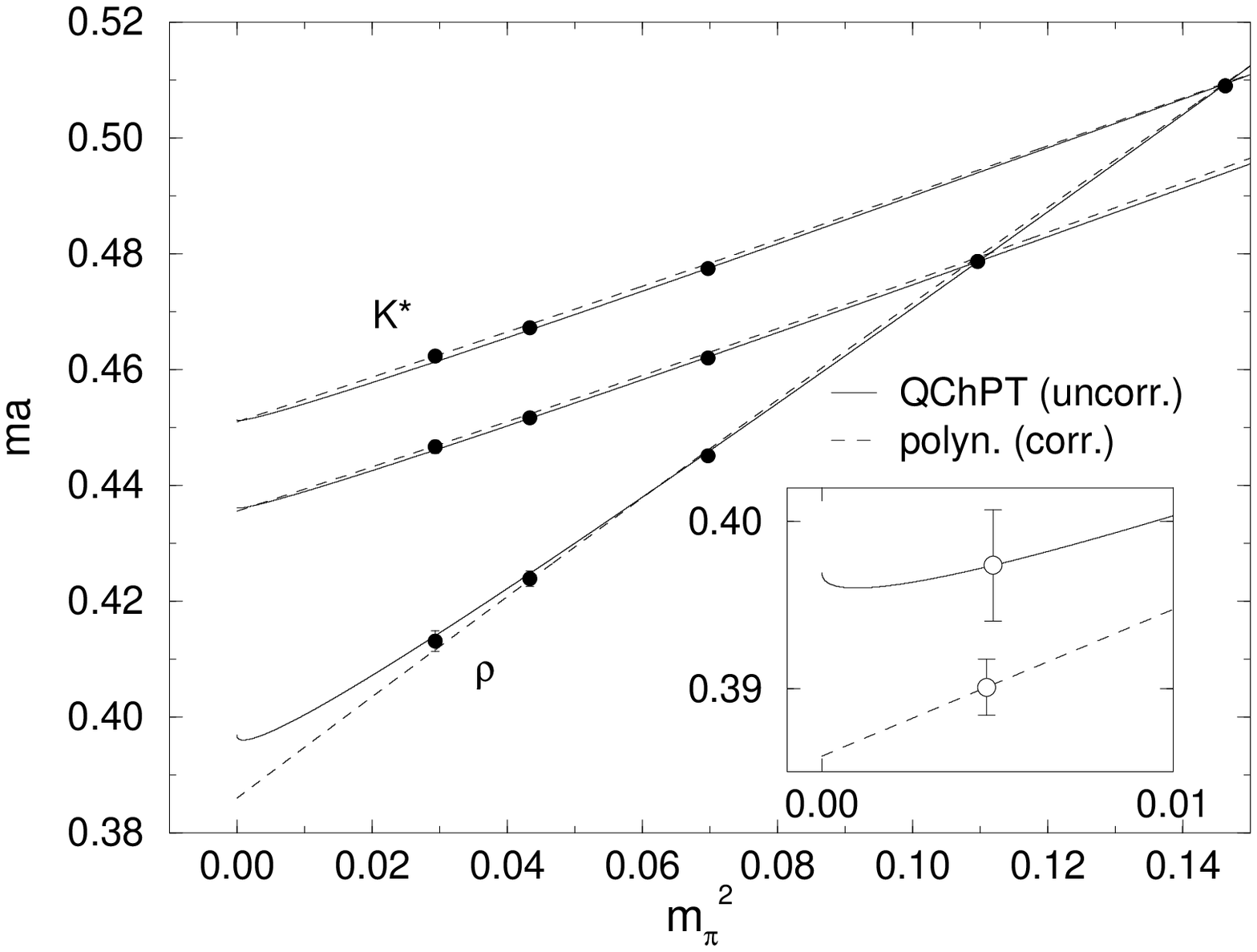}}   
\caption[]{\label{fig:v_chl_quench}
Chiral extrapolations of vector meson masses 
as a function of $m_\pi^2$ at $\beta=5.9$.
Open symbols in the inset represent extrapolated values
at the physical degenerate up and down quark mass.}
\end{minipage}
\hspace{\fill}
\begin{minipage}[b]{80mm}
\centerline{\epsfxsize 73mm \epsfbox{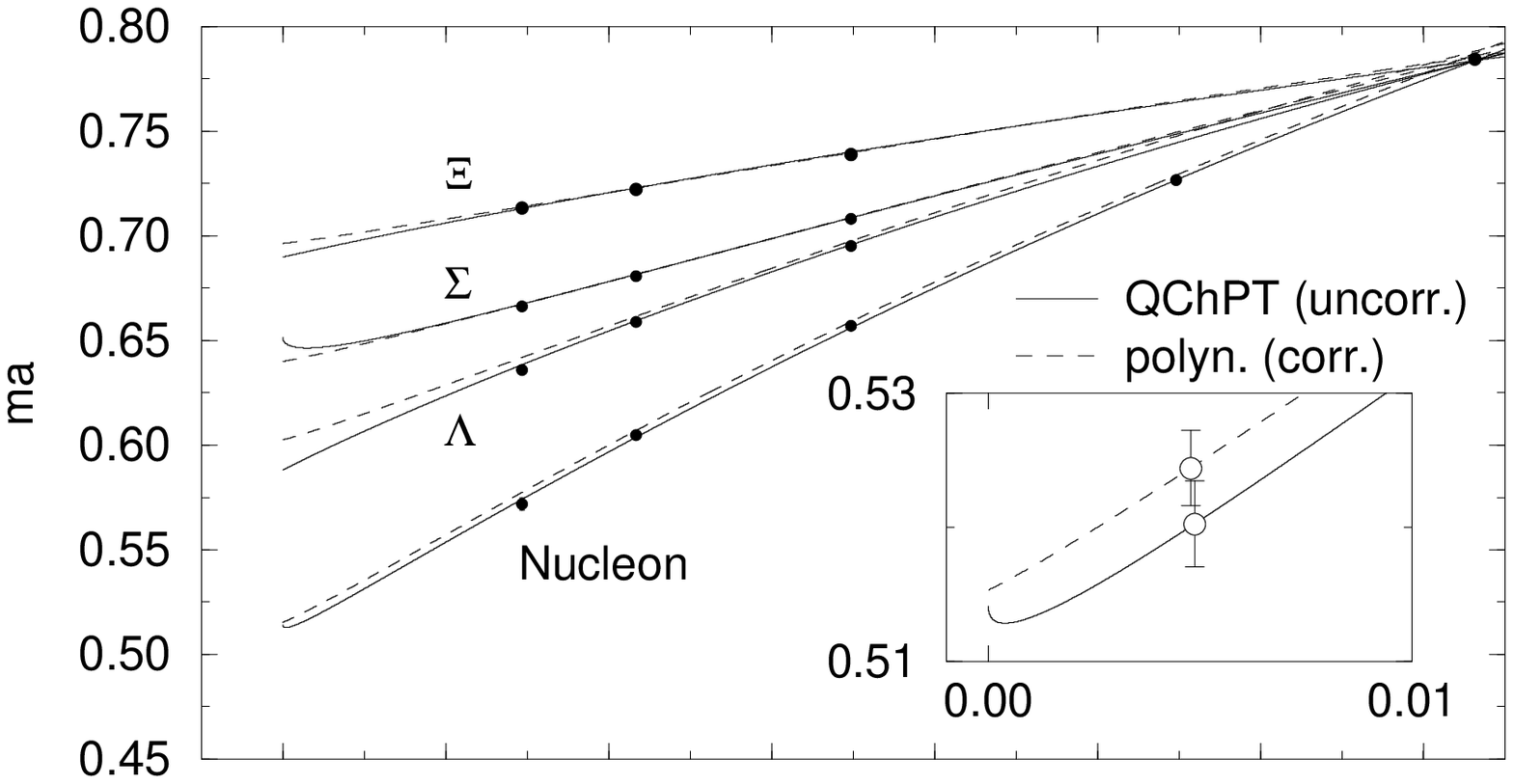}}   
\centerline{\epsfxsize 73mm \epsfbox{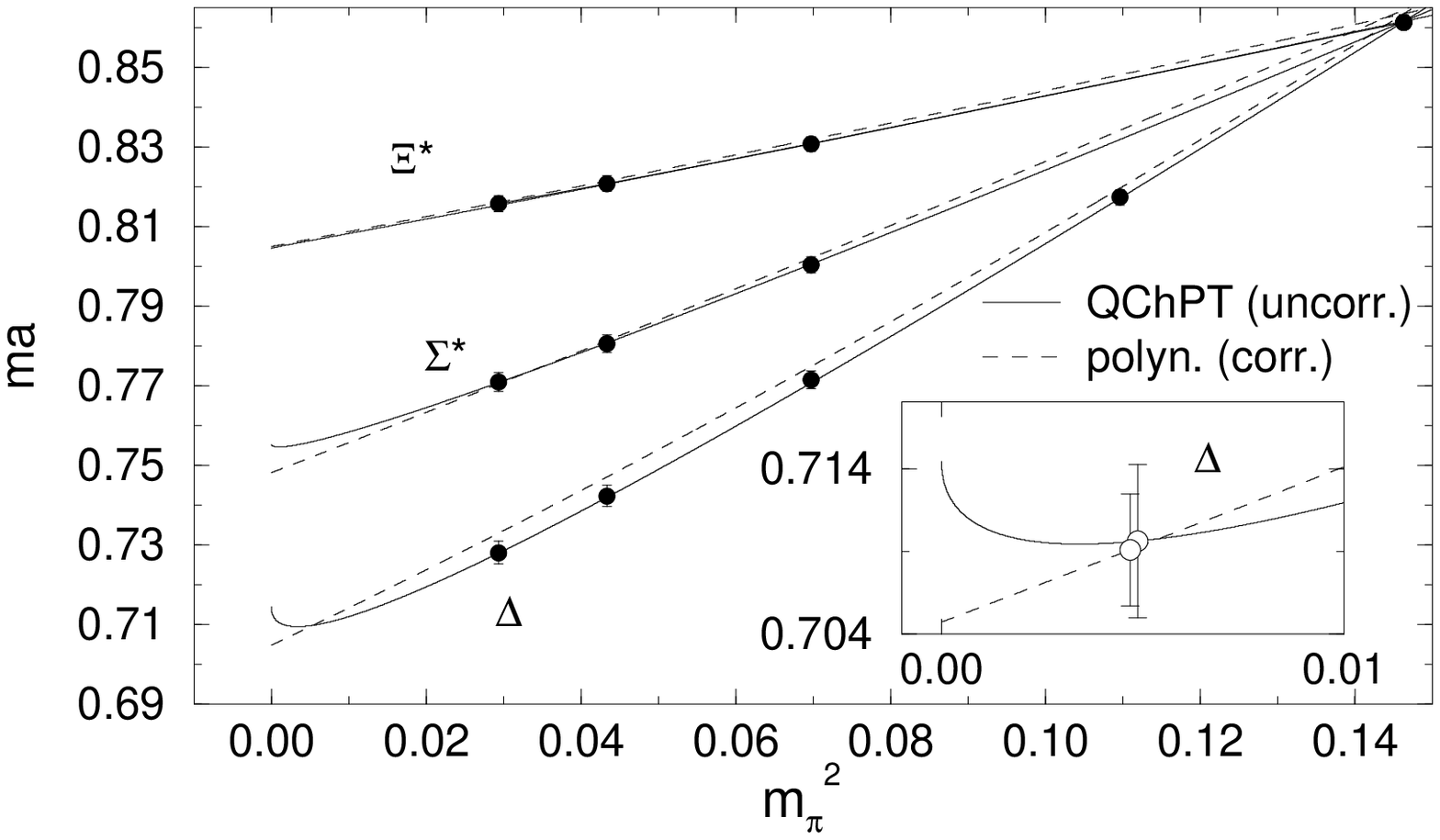}}   
\caption[]{\label{fig:b_chl_quench}
Same as Fig.~\protect{\ref{fig:v_chl_quench}} for octet
and decuplet baryon masses.}
\end{minipage}
\end{figure}

For the vector meson masses we make simultaneous but uncorrelated
fits of the degenerate and non-degenerate data employing
the functional form predicted by QChPT:
\ben
m_{V}&=&m_V^0+\frac{C_{1/2}}{6}
       \left\{\frac{3}{2}(m_\pi+m_\eta)
       +2\frac{m_\eta^3-m_\pi^3}{m_\eta^2-m_\pi^2}\right\} 
       +\frac{C_1}{2}(m_\pi^2+m_\eta^2)
       +C_D(m_\pi^3+m_\eta^3)+C_N m_K^3,
\label{eq:m_vfit}
\een
where $C_{1/2}=-4\pi g_2^2\delta$ with $g_2$ a phenomenological
coupling constant of the vector meson quenched chiral Lagrangian.
It should be noticed that the $O(m_{PS})$ term is 
a characteristic of QChPT. 
This model function up to the $O(m_{PS}^2)$ terms 
describes our vector meson mass data well
as illustrated in Fig.~\ref{fig:v_chl_quench}.
A small bending toward the chiral limit found in the fitting result of
$m_\rho$ reflects the contribution of the $O(m_{PS})$ term.
However, the fitted values for the coefficient $C_{1/2}$ 
of $O(m_{PS})$ term are much smaller than expected.
The average value of $C_{1/2}$ over four $\beta$ points gives 
$C_{1/2}=-0.071(8)$, whose magnitude is ten times smaller than the 
phenomenological estimate $C_{1/2}=-4\pi g_2^2\delta\approx -0.71$
with $\delta=0.1$ and $g_2=0.75$. 
The magnitude of the $O(m_{PS})$ term in eq.(\ref{eq:m_vfit})
is $0.07\times m_\pi\approx 10$MeV, which means 
about a $1\%$ contribution to $m_\rho$.
If we employ the fitting function up to the $O(m_{PS}^3)$ terms, 
we find a few times larger value for $C_{1/2}$ at each beta. 
In this case, however, the fitting results are very unstable
against the $\chi^2$ value. 

\begin{figure}[b]
\begin{minipage}[b]{80mm}
\centerline{\epsfxsize 73mm \epsfbox{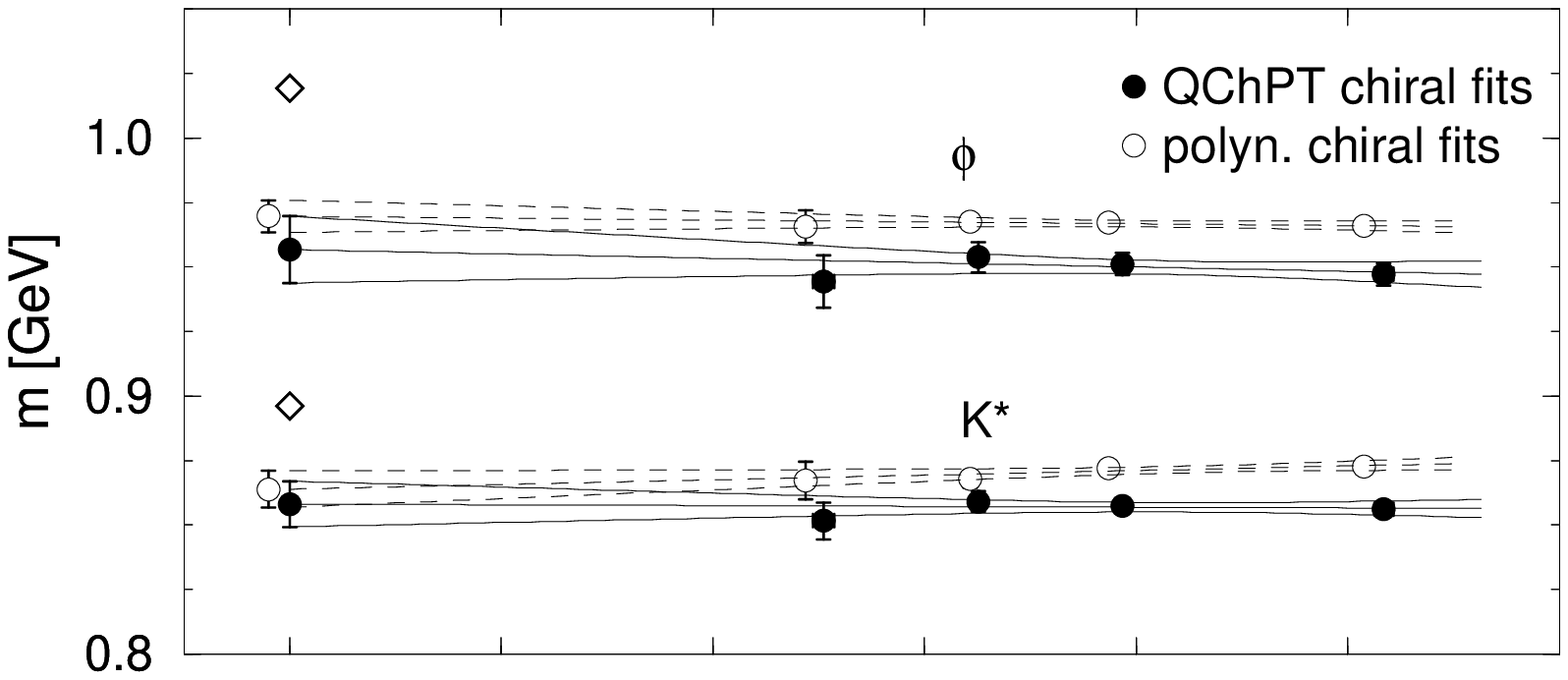}}   
\centerline{\epsfxsize 73mm \epsfbox{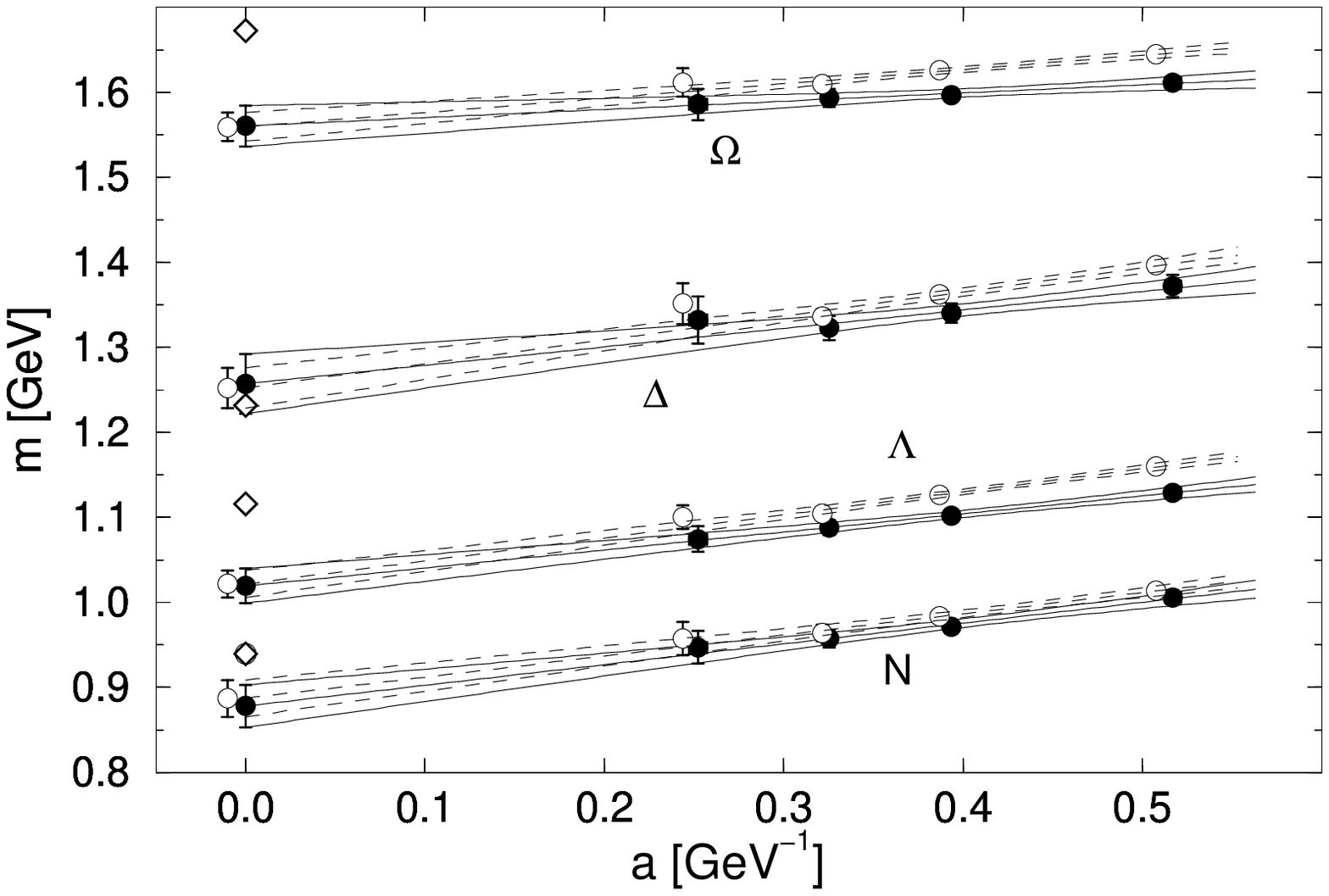}}   
\caption[]{\label{fig:vb_con_quench}
Continuum extrapolations of quenched hadron masses with $m_K$ as 
input for the strange quark mass.}
\end{minipage}
\hspace{\fill}
\begin{minipage}[b]{80mm}
\centerline{\epsfxsize 73mm \epsfbox{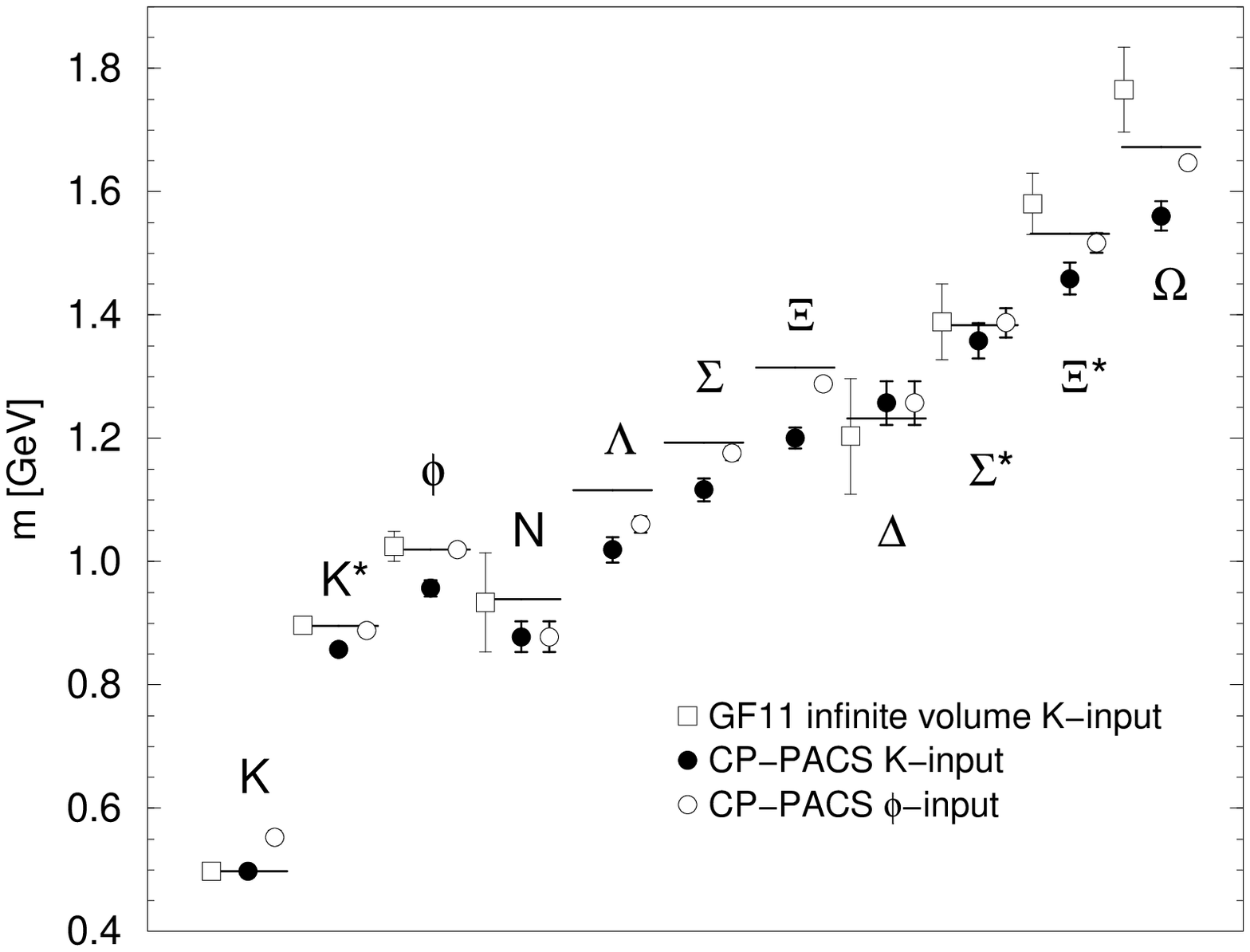}}      
\vskip +.3 cm
\caption[]{\label{fig:spectrum_quench}
Quenched light hadron spectrum in the continuum limit.
GF11 results\protect{\cite{GF11}} are also plotted
for comparison. Horizontal bars denote experimental values.}
\end{minipage}
\end{figure}

For fitting of the octet and decuplet baryon masses
we take the same strategy as for the vector meson mass case. 
Assuming the QChPT mass formulae 
we fit the degenerate and non-degenerate data 
simultaneously without including correlations.
While we find that $O(m_{PS})$ and 
$O(m_{PS}^2)$ terms are sufficient to reproduce 
our decuplet baryon mass data as shown in Fig.~\ref{fig:b_chl_quench}, 
for the octet baryon masses a $O(m_{PS}^3)$ term is inevitably required
to explain the convex curvature of the nucleon mass data.  
The fitted value for the coefficient  $C_{1/2}$ of 
the $O(m_{PS})$ term
in the nucleon is $C_{1/2}=-0.118(4)$ as an averaged  
value over four $\beta$ points.
This is less than half of the phenomenological estimate
$C_{1/2}=-(3\pi/2)(D-3F)^2\delta\approx -0.27$ with
$\delta=0.1$, $F=0.5$ and $D=0.75$.
The $O(m_{PS})$ term contribution to the nucleon mass
is about $16$MeV or $2\%$. 
We find a similar value $C_{1/2}=-0.14(1)$
for $\Delta$.
If the $O(m_{PS}^3)$ terms are included in the fitting function
of the decuplet baryon masses, 
we obtain much smaller value for $C_{1/2}$ in $\Delta$. 

Whereas we observe that the mass formulae based 
on the QChPT reproduce our data adequately,
it is instructive to try another chiral ansatz 
employing polynomial fitting functions in $1/K$ (cubic for $N$, 
quadratic for others). We employ independent fits
for the different degeneracies in the vector meson and the octet
and decuplet baryons. 
The fitting results are shown by dashed lines 
in Figs.~\ref{fig:ps_chl_quench}, \ref{fig:v_chl_quench} 
and \ref{fig:b_chl_quench}. 
Although the QChPT fits seem to give better descriptions
of our hadron mass data than the polynomial ones,
the differences between them are fairly small.  

\subsection{Quenched light hadron spectrum in the continuum limit}

The lattice spacing dependences of the hadron masses are
shown in Fig.~\ref{fig:vb_con_quench}. We expect that the leading 
scaling violation effects are of $O(a)$ in the Wilson quark case.
A linear extrapolation to the
continuum limit assuming the fitting function $m=m_0(1+\alpha a)$
yields $\alpha \sim 0.2$GeV, for which we find that higher order terms
are safely negligible, {\it e.g.}, $(\alpha a)^2\sim 0.01$ 
at $a=0.5$GeV$^{-1}$. 

While the results of the QChPT fits and the polynomial ones 
differ by about $3\%$($5\sigma$) in the largest case 
after the chiral extrapolation at each $\beta$,
the differences in the continuum limit are 
within $1.5\%$($1.5\sigma$) of the results of the QChPT fits.
At a few percent level of statistical errors it is hard to appreciate
the differences between the QChPT chiral extrapolations
and the polynomial ones.

Our final result of the quenched light hadron spectrum is presented 
in Fig.~\ref{fig:spectrum_quench}. The spectrum deviates from experiment
systematically and unambiguously. 
The following discrepancies should be noticed:
The $K$-$K^*$ meson hyperfine
splitting is definitely underestimated by $9.5\%$($4.3\sigma$) for
the $m_K$ input case and by $16\%$($6.1\sigma$) for the $m_\phi$ input case;   
The nucleon mass is appreciably smaller than experiment 
by $7\%$($2.5\sigma$);
For the strange octet baryon masses our data 
with the $m_K$ input are systematically
smaller than experiment by $6-9\%$($4-7\sigma$), 
which is much reduced in the $m_\phi$ input case;
While the $m_K$ input leads $30\%$ smaller estimates for
the decuplet mass splittings in average, we observe rather 
good agreement for the $m_\phi$ input results.  
 
\section{Light hadron spectrum in two-flavor full QCD}

\subsection{Details of numerical simulation}

\nopagebreak

The simulation of full QCD requires a huge amount of 
computing time compared to the quenched approximation.
In order to secure an adequate physical spatial lattice size
for avoiding finite size effects, the lattice spacing
is compelled to be coarse $a^{-1}\simlt 2$GeV, which
urges us to employ improved actions.
Our choice is a RG-improved gauge action\cite{RGaction} 
and the SW quark action\cite{SWaction} with a mean-field improved value
of $c_{SW}$. This decision is based on a former comparative 
study of various combinations of improved gauge and quark 
actions at $a^{-1}\approx 1$GeV\cite{prestudy}.
  
Our full QCD study is performed with two flavors of sea quarks
which are supposed to be the degenerate dynamical up and down quarks.
Our simulation parameters are summarized in Table~\ref{tab:para_full}.
Four $\beta$ values covering the lattice spacing 
in the range $a^{-1}\approx 1-2$GeV are employed to
examine the scaling behavior of the light hadron spectrum. 
We keep the physical spatial lattice size approximately constant
at $La\approx 2.4$fm except for the finest lattice where
$La\approx 2.0$fm.
At each value of $\beta$ we choose four values of sea quark mass
corresponding to $m_\pi/m_\rho\approx 0.8$, $0.75$, $0.7$ and $0.6$.
Gauge configurations including the dynamical sea quark effects
are generated with the HMC algorithm.
  
We calculate hadron masses
using the five valence quarks whose masses correspond to
$m_\pi/m_\rho\approx 0.8$, $0.75$, $0.7$, $0.6$ and $0.5$.
Hadron propagators are constructed with 
the smeared source quark propagators 
at every fifth trajectory. Hadron masses are extracted by
an exponential fit ignoring correlations between time slices. 
Errors are estimated by the jackknife
method with a bin size of 10 configurations (50 HMC trajectories).

The light hadron spectrum is obtained by setting the
sea quark mass on the physical up and down quark mass
and  the valence quark mass on the physical up 
and down quark mass or the strange one.
The physical point of the degenerate up 
and down quark mass is fixed with $m_\pi$ and $m_\rho$ as input.
For the strange quark mass the physical point is determined from
$m_K$ or $m_\phi$.
The lattice scale $a^{-1}$ is set with $m_\rho$.  

\subsection{Chiral extrapolations in the sea and valence quark masses}

\begin{figure}[b]
\begin{minipage}[b]{80mm}
\centerline{\epsfxsize 68.5mm \epsfbox{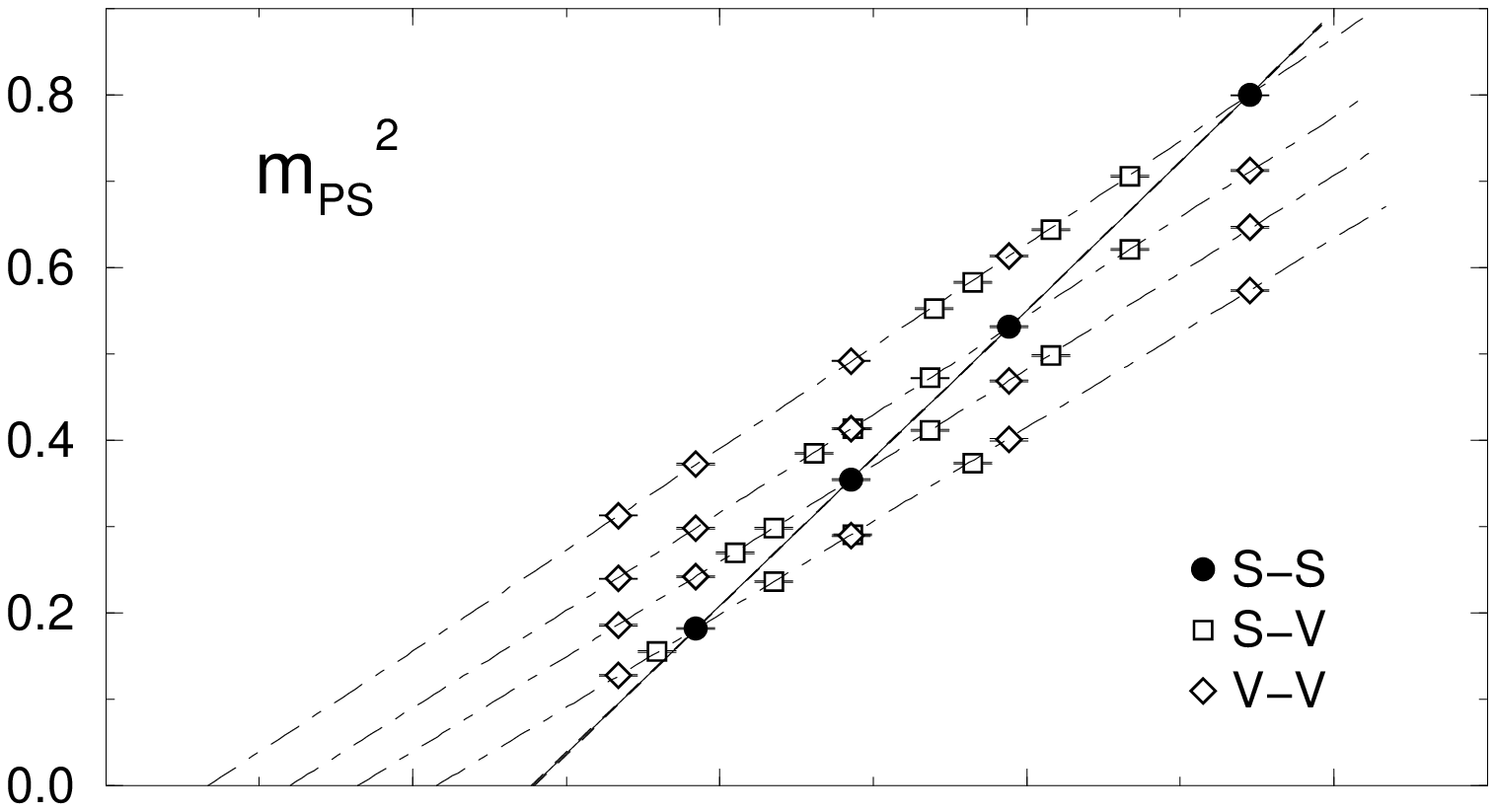}}   
\centerline{\epsfxsize 68.5mm \epsfbox{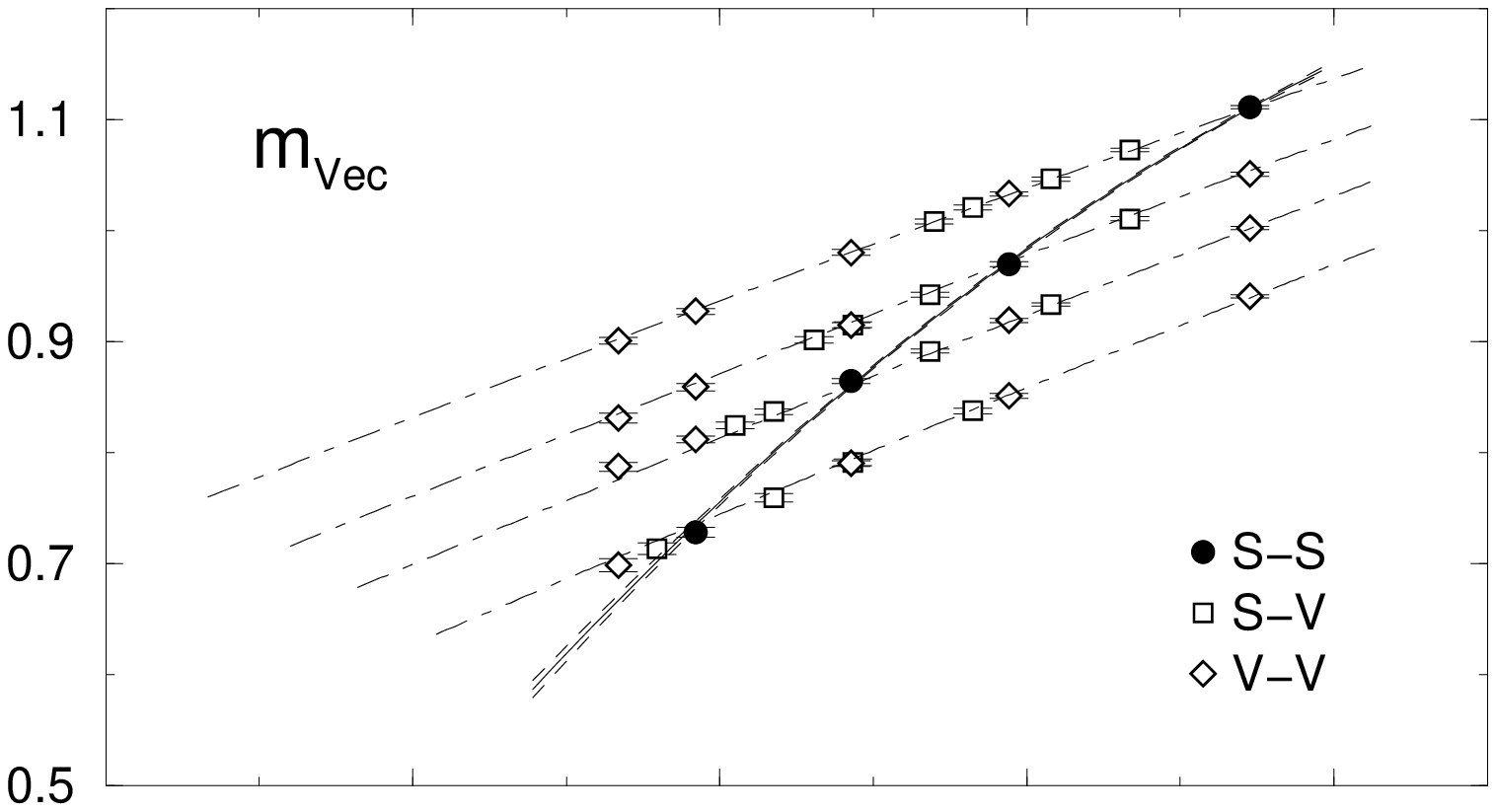}}   
\centerline{\epsfxsize 68.5mm \epsfbox{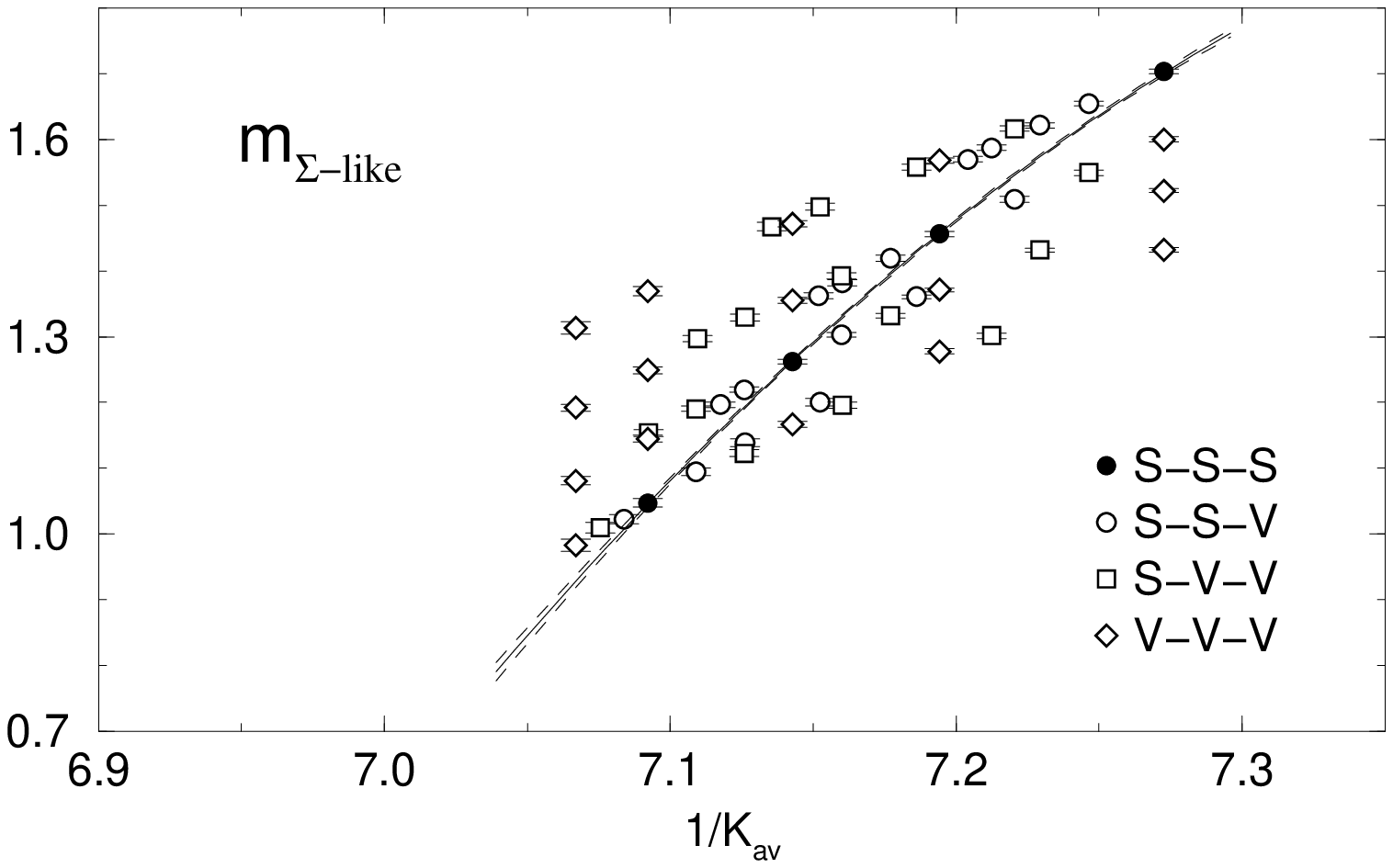}}   
\caption[]{\label{fig:msn_chl_full}
Sea and valence quark mass dependence of hadron masses
at $\beta=1.95$. Solid and broken lines represent
results of quadratic fits for chiral extrapolations.
See text for the labels $S$ and $V$.}
\end{minipage}
\hspace{\fill}
\begin{minipage}[b]{80mm}
\centerline{\epsfxsize 73mm \epsfbox{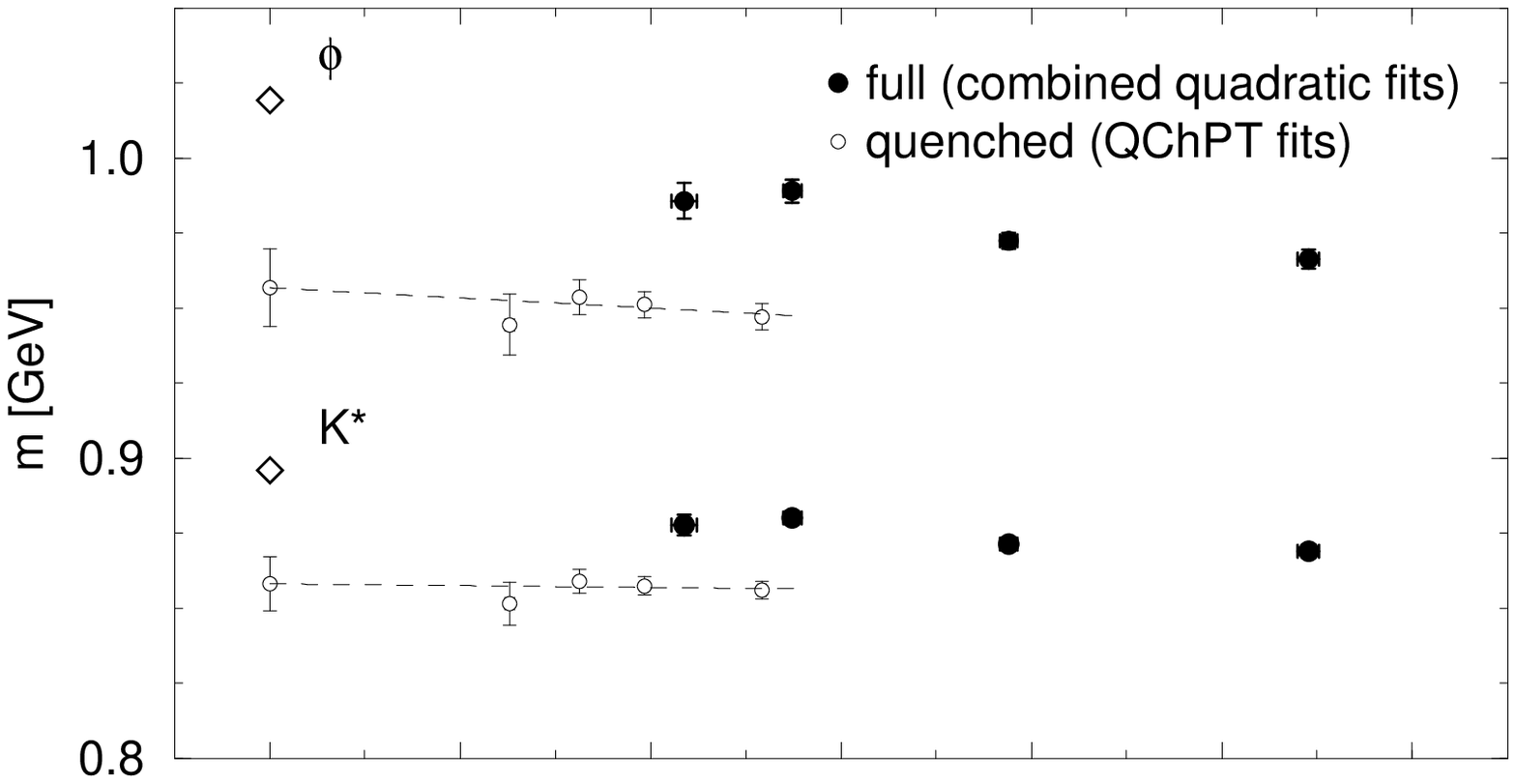}}   
\centerline{\epsfxsize 73mm \epsfbox{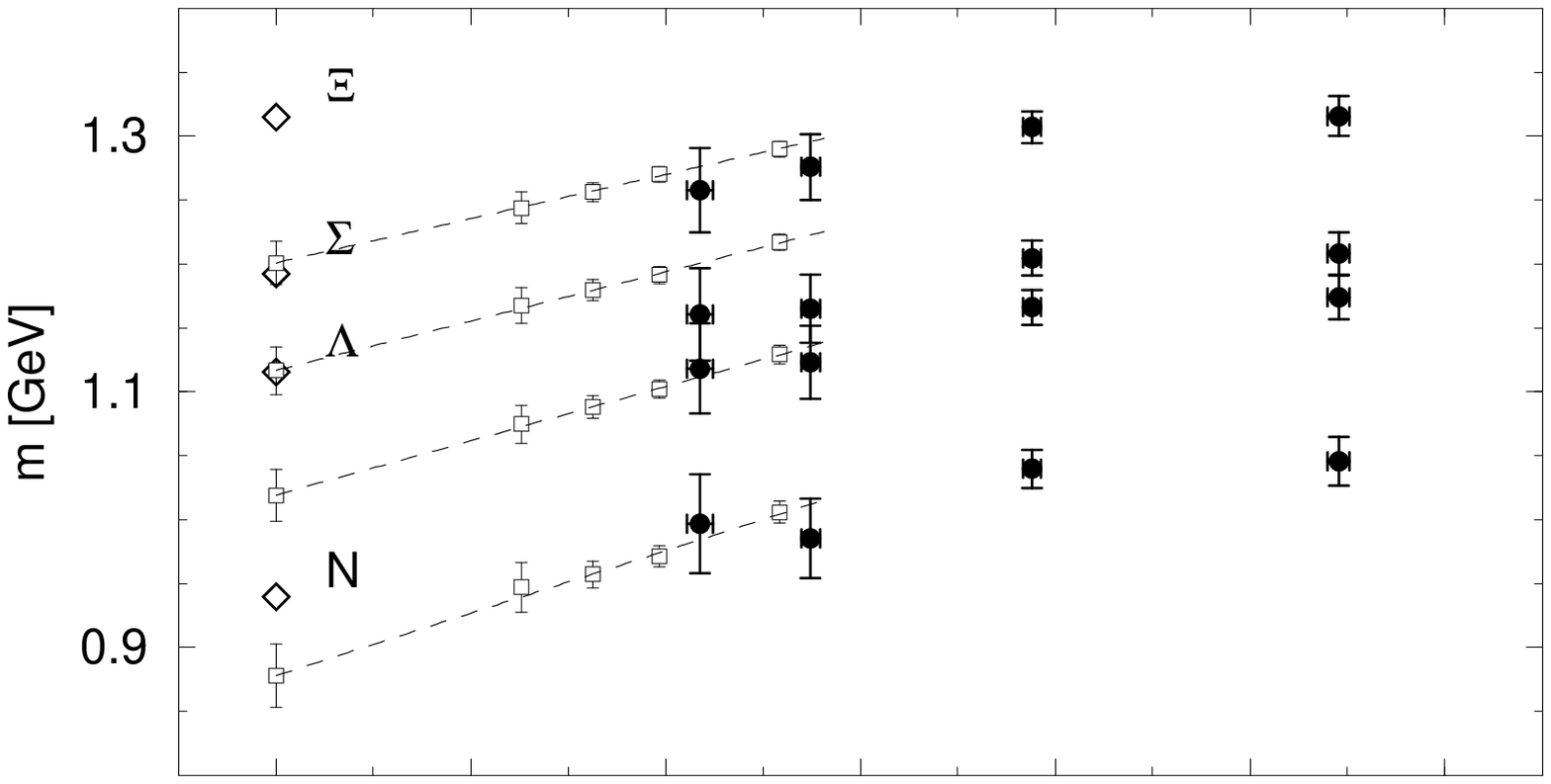}}   
\vspace*{1mm}
\centerline{\epsfxsize 73mm \epsfbox{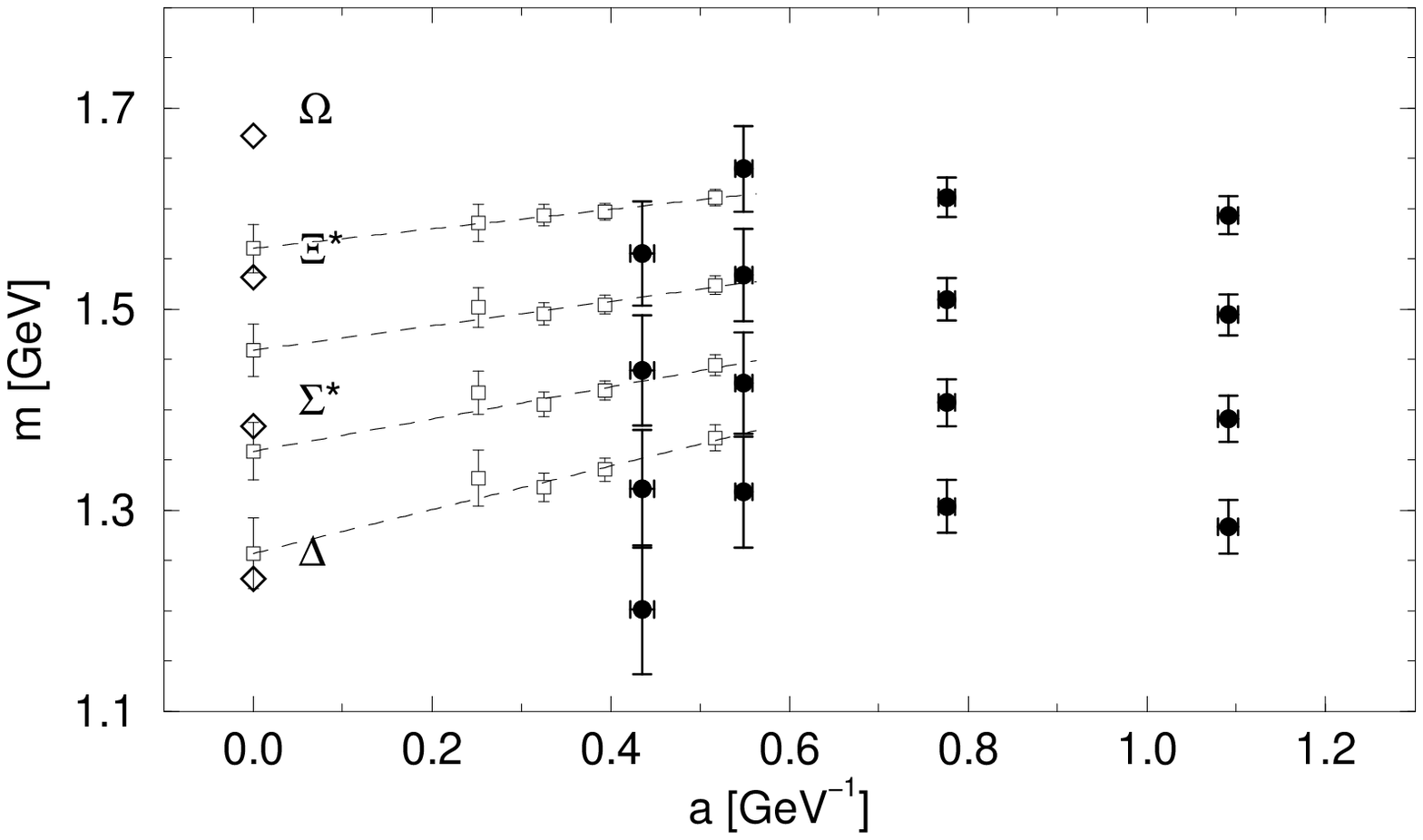}}   
\caption[]{\label{fig:spectrum_full}
Light hadron spectrum in two-flavor full QCD
with $m_K$ as input for the strange quark mass
as a function of the lattice spacing.
For comparison quenched QCD results in the last section
are also plotted.}
\end{minipage}
\end{figure}

In Fig.~\ref{fig:msn_chl_full} we plot a result
for the sea and valence quark mass dependences of 
hadron masses at $\beta=1.95$. The data are parameterized 
by the averaged hopping parameter 
$1/K_{av}=(1/K_{val(1)}+1/K_{val(2)})/2$ of the two kind of 
valence quarks constituting the hadrons. 
$S$ represents a valence quark with $K_{val}=K_{sea}$ and
$V$ a valence quark with $K_{val}\ne K_{sea}$.

\begin{table}[t]
\caption{\label{tab:para_full}
Simulation parameters for full QCD.}
\begin{tabular}{lccccllll}
$\beta$ & $L^3\times T$ & $c_{SW}$ & $a$[fm] & $La$[fm] 
& \multicolumn{4}{c}{$m_{\pi}/m_{\rho}$ for sea quarks} \\
        &               &          &         &
& \multicolumn{4}{c}{\#traj.} \\
\tableline 
1.80 & $12^3\times 24$ & $1.60$ & 0.215(2) & 2.58(3) 
& 0.8060(7) & 0.753(1)  & 0.696(2) & 0.548(4) \\ 
&&&&& 6250  & 5000      & 7000     & 5250 \\
1.95 & $16^3\times 32$ & $1.53$ & 0.153(2) & 2.45(3) 
& 0.8048(9) & 0.751(1)  & 0.688(1) & 0.586(3) \\ 
&&&&& 7000  & 7000      & 7000     & 5000 \\ 
2.10 & $24^3\times 48$ & $1.47$ & 0.108(2) & 2.59(5) 
& 0.806(2)  & 0.757(2)  & 0.690(3) & 0.575(6) \\ 
&&&&& 2000  & 2000      & 2000     & 2000 \\
2.20 & $24^3\times 48$ & $1.44$ & 0.086(3) & 2.06(6) 
& 0.800(2)  & 0.754(2)  & 0.704(3) & 0.629(5) \\  
&&&&& 2000  & 2000      & 2000     & 2000 \\
\end{tabular}
\end{table}

We observe that the data of the meson masses 
are distributed on different four lines
in accordance with the four sea quarks.
Partially quenched data on each sea quark
show almost linear behavior both for $m_{PS}^2$ and $m_{Vec}$. 
Their slopes, however, slightly depend on the
sea quark masses: As the sea quark mass decreases 
the slope decreases for $m_{PS}^2$, 
while the slope increases for $m_{Vec}$.
In the $S$-$S$ channel 
$m_{PS}^2$ behaves almost linearly in 
$1/K_{sea}$, whereas $m_{Vec}$ exhibit a convex curvature.
Chiral extrapolations of the meson masses  
are made by global fits of all the data
assuming quadratic functions 
in terms of $1/K_{sea}$ and $1/K_{val}$.
For the PS meson mass we employ
\ben
m_{PS}^2&=& B_s m_{sea}+B_v{\overline m}_{val}+C_s m_{sea}^2
+C_v{\overline m}_{val}^2+C_{sv}m_{sea}{\overline m}_{val}
+C_{12}m_{val(1)}m_{val(2)},
\een
where bare quark masses are defined by 
$m_{sea/val(i)}=(1/K_{sea/val(i)}-1/K_c)/2$ and 
${\overline m}_{val}$ is the averaged value of two 
valence quark masses.
Similar fitting function without the valence-valence cross term
$m_{val(1)}m_{val(2)}$ is used for
the vector meson masses and the decuplet baryon masses. 
The fitting results for the meson masses are drawn 
in Fig.~\ref{fig:msn_chl_full}.

As for the octet baryon masses we find a rather complicated
situation in Fig.~\ref{fig:msn_chl_full}: Partially quenched 
baryon masses are not functions of ${\overline m}_{val}$. 
For the fitting function we take a combination of linear terms
based on the ChPT prediction and general quadratic terms of 
individual $m_{val(i)}$. This function has 12 free parameters 
in all, which are determined by a combined 
fit of $\Sigma$-like and $\Lambda$-like baryon
masses. 

\subsection{Sea quark effects on the light hadron spectrum}

The results for the light hadron spectrum are presented
in Fig.~\ref{fig:spectrum_full}
as a function of the lattice spacing. For comparative purpose
we also plot the quenched results(open symbols) 
from the last section, which would help us distinguish 
the dynamical sea quark effects.

We observe an intriguing feature in the meson sector:
The $K^*$ and $\phi$ meson masses in full QCD are heavier than
those in quenched QCD at finite lattice spacing and become closer
to experimental values as the lattice spacing decreases.
However, our full QCD results do not necessarily
agree with experiment in the continuum limit.
We have two possible reasons.
One is the relatively heavy sea quark masses corresponding
to $m_\pi/m_\rho\approx 0.6-0.8$, for which 
the chiral extrapolations are rather ambiguous.  
The other is the quenched treatment of the strange quark.

For the baryon sector it is hard to read any meaningful
implication in comparison between the full QCD results 
and the quenched ones. At the finite lattice spacing
the full QCD spectrum is consistent with the quenched ones
within rather large errors, while the lattice spacing dependence
of the decuplet baryon masses in full QCD 
is opposite to that in quenched case.    

\section{Conclusions}

We have presented our results for the quenched
light hadron spectrum. After examining the validity of
QChPT in the pseudoscalar meson masses
we applied its hadron mass formulae  
to the chiral extrapolations. 
Although the QChPT predictions reproduce
our hadron mass data well, 
for the vector meson and baryon masses
it is hard to confirm the contributions of the term
specific to QChPT. 
The spectrum in the continuum limit deviates 
systematically and unambiguously from experiment
on a level of about $10\%$ far beyond the statistical error
of $1-2\%$ for mesons and $2-3\%$ for baryons.

Our full QCD study is an exploring step 
toward a realistic QCD simulation.
For the light hadron spectrum we found an 
encouraging result that the deficiency of 
the meson hyperfine splitting in the quenched approximation
is largely compensated with the dynamical sea quark effects.
At this stage, however, the most important conclusion
is as follows: 
To perform a close investigation of the sea quark effects 
we need a direct comparison between the full QCD results and
the quenched ones employing the same gauge and quark actions
with the same $a^{-1}$, $m_\pi/m_\rho$ and $La$.
Work in this direction is now in progress.

I am grateful to all the members of the CP-PACS Collaboration for
their help in preparing this manuscript.
I would particularly like to thank R.~Burkhalter and T.~Yoshi{\'e}.
This work is supported in part by the Grants-in-Aid of Ministry
of Education(Nos. 08NP0101 and 09304029).


\begin{references}  

\bibitem{GF11}F.~Butler {\it et al.}, Nucl. Phys. {\bf B430}, 179 (1994).

\bibitem{latt98}CP-PACS Collaboration, 
S.~Aoki {\it et al.}, hep-lat/9809146;
R.~Burkhalter, hep-lat/9810043.

\bibitem{finiteV}MILC Collaboration, C.~Bernard {\it et al.},
Nucl. Phys. B (Proc. Suppl.) {\bf 60A}, 3 (1998).

\bibitem{QChPT}C.~W.~Bernard and M.~F.~L.~Golterman, 
Phys. Rev. {\bf D46}, 853 (1992);
S.~R.~Sharpe, Phys. Rev. {\bf D46}, 3146 (1992);
J.~N.~Labrenz and S.~R.~Sharpe, Phys. Rev. {\bf D54}, 4595 (1996)
M.~Booth {\it et al.}, Phys. Rev. {\bf D55}, 3092 (1997).

\bibitem{m_q^AWI}M.~Bochicchio {\it et al.},  
Nucl. Phys. {\bf B262}, 331 (1985).


\bibitem{RGaction}Y.~Iwasaki, Nucl. Phys. {\bf B258}, 141 (1985).

\bibitem{SWaction}B.~Sheikholeslami and R.~Wohlert, 
Nucl. Phys. {\bf B259}, 572 (1985).

\bibitem{prestudy}CP-PACS Collaboration, 
S.~Aoki {\it et al.}, hep-lat/9902018.

\end{references}
\end{document}